\documentclass[aps,showpacs,reprint,pre]{revtex4-1}

\usepackage{graphicx}
\usepackage{bm}
\usepackage{CJK}

\begin{document}
\begin{CJK*}{UTF8}{gbsn}
\title{Monte Carlo Algorithm for Simulating Reversible Aggregation of Multisite Particles}
\author{Qiang Chang (常强)}
\author{Jin Yang (杨劲)}
\email{jinyang2004@gmail.com}
\affiliation{Chinese Academy of Sciences-Max Plank Society Partner Institute for Computational Biology \\ Shanghai Institutes for Biological Sciences, Shanghai 200031, China}

\begin{abstract}
We present an efficient and exact Monte Carlo algorithm to simulate reversible aggregation of particles with dedicated binding sites. This method introduces a novel data structure of dynamic bond tree to record clusters and sequences of bond formations. The algorithm achieves a constant time cost for processing cluster association and a cost between $\mathcal{O}(\log M)$ and $\mathcal{O}(M)$ for processing bond dissociation in clusters with $M$ bonds. The algorithm is statistically exact and can reproduce results obtained by the standard method. We applied the method to simulate a trivalent ligand and a bivalent receptor clustering system and obtained an average scaling of $\mathcal{O}(M^{0.45})$ for processing bond dissociation in acyclic aggregation, compared to a linear scaling with the cluster size in standard methods. The algorithm also demands substantially less memory than the conventional method.
\end{abstract}


\maketitle

\end{CJK*}
\section{Introduction}\label{sec:intro}

Reversible aggregation or self-assembly of particles with multiple interactive sites is of fundamental importance to diverse processes in physical and living systems including aggregation of colloidal particles~\cite{lin1989universality} and proteins~\cite{fields1992theory}, synthesis of supramolecules in polymer science~\cite{cordier2008self}, and self-assembly of patchy particles such as nanoparticles~\cite{mirkin1996dna,hermans2009self} and synthetic biomolecules~\cite{bilgicer2007synthetic} in material sciences~\cite{glotzer2007anisotropy,*pawar2010fabrication}. Reversible aggregation was traditionally studied using the generalized Smoluchowski equation~\cite{family1986kinetics,odriozola2002constant} that requires one to develop kernel functions for cluster aggregation and fragmentation to obtain the kinetics of the cluster size distribution. Proper kernel functions can be analytically characterized often under restrictive assumptions of particle interactions. For acyclic aggregation of multisite particles that forms loopless clusters, Wertheim's thermodynamic perturbation theory~\cite{wertheim1984fluidsi, *wertheim1984fluidsii,*wertheim1986fluidsiii,*wertheim1986fluidsiv} and Flory-Stockmayer theory~\cite{flory1953principles} can predict equilibrium properties for simple systems. To study more general systems, Monte Carlo simulations are indispensable to provide new insights into the kinetics and equilibrium properties of the aggregation.

\begin{figure}[t]
\centering
\includegraphics[scale=0.37]{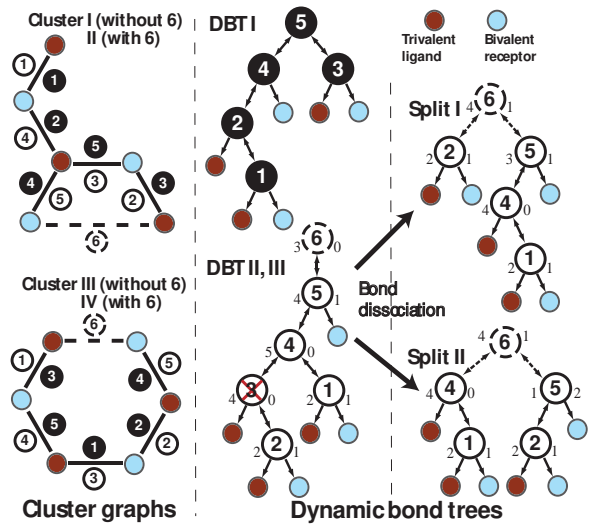}
\caption{\label{fig:struct} (Color online) A trivalent ligand and bivalent receptor (TLBR) aggregation system. Left panel: cluster connectivity graphs with nodes as particles and edges as interparticle bonds.  Acyclic clusters I and III (without bond 6) and cyclic II and IV (with bond 6) have the same number of ligands and receptors but different topologies (not drawn to reflect the actual biochemical structure of a cluster). Middle panel: acyclic clusters I and III are represented by either DBT I or DBT II (without bond 6), depending on the sequence of bond formation. Bonds in cluster graphs and in corresponding DBTs are labelled numerically (with filled or unfilled circles), reflecting the order of bond formation. DBT III (with bond 6) represents cyclic cluster II or IV. Right panel: two scenarios of cluster dissociation (Split I and II) after breaking bond 3 (in DBT II or III). The number of free ligand sites (left) and the number of free receptor sites are shown beside individual a DBT node as weights for the subcluster represented by the individual DBT node.}
\end{figure}

Reversible aggregation involves two principal types of reaction processes, bond formation and breaking. The balance of these two competing processes allows an aggregation system to reach an equilibrium after a transient phase. In the standard site-based simulation algorithm, clusters are stored as graphs representing the connectivity between particles (see Fig.~\ref{fig:struct} left panel for an example of multivalent ligand-receptor interaction system). To resolve information such as composition and topology of clusters, graph traversals by depth-first (or breadth-first) search are routinely applied, which are often computationally costly.

To simulate irreversible aggregation that ignores bond breaking, a highly efficient algorithm, the classic {\it weighted union-find with path compression}~\cite{IntroAlg2001}, can identify cluster membership of binding sites and amalgamate two clusters in a near constant time proved by Tarjan~\cite{tarjan1975} and demonstrated in computing site or bond percolation models~\cite{newman2000efficient}. The algorithm employs a tree-based data structure to index the cluster membership of individual sites. However, this strategy cannot be readily adopted to simulate reversible aggregation because bond dissociation requires time-consuming reorganization of the tree-based data structures used in the algorithm. 

To simulate reversible aggregation, the standard site-based algorithm labels each individual site to identify its cluster membership. Bond formation and dissociation require site relabelings whenever a reactive event is sampled. In an event of bond formation between two sites, one first determines whether both sites belong to a same cluster by comparing their labels. The two sites belong to two separate clusters if the labels are different. In the latter case, one needs to relabel all sites in one cluster with the label of the other, which is done by a graph traversal of one cluster that is to be relabeled. Because a cluster size is known by simple bookkeeping, one can always relabel sites in the smaller cluster with the label assigned to the larger one to minimize the cost. This heuristics usually improves performance to a substantial extent over relabeling an arbitrary subcluster. Unfortunately, this weighted relabeling is infeasible for processing dissociation of a cluster into two smaller ones because the sizes of the two resulting subclusters are not known {\it a priori} (bookkeeping such information is nontrivial and expensive). Therefore, upon identification of a bond to break, by graph traversal one systematically relabels an arbitrary subcluster to which the bond connects. For a cyclic cluster with loops, a graph traversal also identifies whether the dissociating bond resides in a loop in order to decide whether site relabeling is needed. The average time complexity of a cluster traversal is $\mathcal{O}(N+M)$, scaled by the size of the cluster, here measured as the number of particles $N$ and the number of bonds $M$ in the cluster. Clearly, the standard algorithm becomes computationally intensive in particular for simulating high density systems that contain giant clusters recorded by large connectivity graphs.

Here, we present an efficient kinetic Monte Carlo algorithm that amalgamates two clusters in $\mathcal{O}(N_C)$ time, where $N_C$ is the average number of aggregates, and the algorithm splits a cluster in time between $\mathcal{O}(\log{M})$ and $\mathcal{O}(M)$. Unlike the site-based methods, the main idea behind our algorithm is based on the observation that explicit cluster graphs are usually not required in a simulation. Instead of using connectivity graphs, we use a more efficient data structure, namely {\em dynamic bond tree} (DBT), to track bonds and clusters without recording and updating the actual connections between particle sites. As a considerable advantage, the algorithm replaces expensive traversals of connectivity graphs with much more efficient updates of DBTs. The algorithm is numerically exact in generating observable quantities such as the cluster size distribution, average cluster size and the number of clusters. The algorithm is directly applicable to simulate aggregations that allow formation of both acyclic and cyclic clusters. If topologies of clusters are of interest, connections among sites can be recorded in parallel during a simulation, or alternatively, ensembles of cluster topologies can be mapped out stochastically from the corresponding DBTs by postprocessing.

The paper is organized as follows. In Section~\ref{sec:alg}, we elaborate the details about the data structure in our algorithm. Using acyclic aggregation as an example, we explain how to compute the two basic events of bond formation and dissociation using the data structure. The complete algorithm is summarized by the end of the section and the adaptation of the algorithm to simulating cyclic aggregation is explained. In Section~\ref{sec:app}, we evaluate the performance of our algorithm by applying it to simulate a multivalent ligand-receptor binding model and compare it to the site-based graph traversal algorithm.

\section{The Algorithm}\label{sec:alg}

{\bf Data structure for storing clusters -- Dynamic bond tree.} To simplify explaining the algorithm without loss of generality, we consider a system that contains a homogeneous population of particles, each of which is decorated with one or more symmetric surface patches (binding sites). We assume that a single binding site can only sustain at most one bond. As demonstrated below in Section III (Application), the algorithm can be readily extended to a system with a heterogeneous population of particles with non-identical sites that can bind to complementary sites on other particles. The basic data structure in our algorithm is the dynamic bond tree (DBT) used to store every cluster of particles. Fig.~\ref{fig:struct} illustrates structures of DBTs for both acyclic and cyclic clusters in an example system of multivalent ligand receptor binding, which will be later used as an application to demonstrate the DBT-based simulation algorithm.

Each multiparticle cluster is identified by the root node of the corresponding DBT. A leaf node in a DBT represents a single particle in the cluster, whereas a non-leaf node including the root node records a site-site bond. Each non-leaf node has either one or two child nodes, depending on whether the bond is formed between intracluster sites or intercluster sites, respectively. A node with two children (e.g., non-leaf nodes in DBTs in Fig.~\ref{fig:struct} except for node 6) indicates that the bond was formed by an association between a pair of sites that reside on two previously separate clusters represented by the two child nodes, whereas a node with a single child (e.g., node 6 in Fig.~\ref{fig:struct} middle panel) indicates that the bond was formed by an association between a pair of sites that reside on a same cluster represented by the only child node. By these conventions, a cluster is {\it cyclic} if and only if the corresponding DBT contains at least one non-leaf node that has a single child. Otherwise, a cluster is {\it acyclic}. Unlike the standard cluster connectivity graph, a DBT does not require keeping track individual binding sites. Below, we use acyclic aggregation as an example to describe how to process bond formation and breaking using DBT structures, and later we explain that processing cyclic aggregation only requires slight adaptation.

{\bf Bond formation (for acyclic clusters).} To process a bond formation, two clusters (note that one or both could be free particles) are first sampled according to their joint probability of contributing binding sites. The probability for a cluster $c$ to contribute a binding site can be related to its number of free sites $s_c$ with a function $g(s_c)$, whose value is assigned to each cluster as a weight (see Fig.~\ref{fig:struct} for example weights assigned to DBT nodes). In a simplest  form, $g(s_c)$ can be usually considered to be proportional to $s_c$, but we note that in general the function $g(s_c)$ might assume different forms in different systems or models. For example, consider that a cluster of a spherical volume has $s_c$ free binding sites. Due to the effect of steric hindrance, one may assume that only free sites near the cluster surface can form a bond with a site near the surface on another cluster. In this model, assuming free sites are homogeneously distributed within the cluster volume and on the surface, one can show that $g(s_c)\sim s_c^{2/3}$ is a good approximation.

After two binding clusters are determined, a new node $z$ is then created as a root node of the DBT that will store the resulting cluster. The root nodes, $x$ and $y$, of the DBTs of the two binding clusters become two children nodes (representing two subclusters) of the root node $z$. A weight of value $g(s_z)$ is then assigned to $z$. The number of free sites in the newly formed cluster is $s_z=s_x+s_y-2$, where the adjustment by $-2$ is due to the consumption of two sites to form bond $z$, each from one subcluster. Obviously, the process of constructing a DBT manifests the hierarchical nature of cluster aggregation, in which a bond node in a lower level in the DBT formed earlier than one in an upper level. In fact, this structure of hierarchy in DBTs underlies the performance gain of our algorithm. Unlike the standard method, this procedure of merging two clusters does not require cluster membership checking of trial binding sites and systematical site relabeling, and thus merely has a constant time complexity. We note that locating two clusters to bind demands searching over the entire array of clusters. Therefore, the overall complexity of bond association scales linearly with the number of clusters (i.e., $\mathcal{O}(N_C)$, where $N_C$ is the total number of clusters). However, we will show in the example TLBR system below that this cost is in most cases modest if it is not ignorable. In particular, when a system reaches the highly aggregated regime, processing bond formation has a near constant time cost because the number of clusters $N_C$ remains small and grows very slowly with the number of particles~\cite{yang2008rejection}.

\begin{figure}[b]
\centering
\includegraphics[scale=0.34]{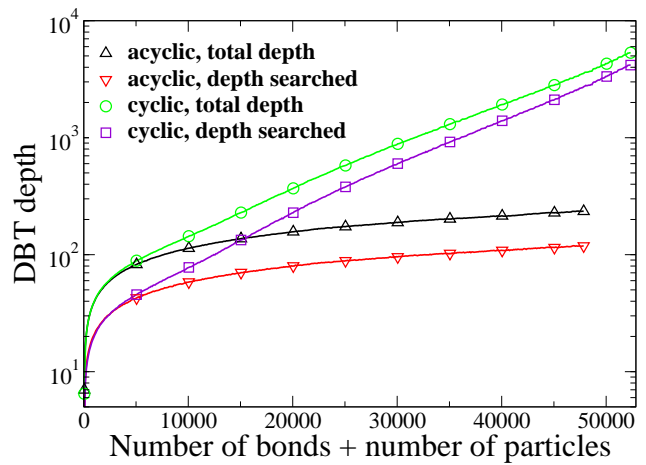}
\caption{\label{fig:dep} (Color online) Cluster size ($N+M$) vs. depth of the corresponding DBT (total and searched), for acyclic or cyclic aggregations. Simulations use $15000$ bivalent receptors and $10000$ trivalent ligands (i.e., total ligand sites and total receptor sites are equal), with fixed rate constants $k_+=6.67\times 10^7 s^{-1}$ and $k_{++}=100k_+$. The dissociation rate constant $k_{\rm off}$ was varied for simulations to generate clusters of different sizes. Cluster size and the DBT depth were obtained by averaging. The data points of cluster size are binned within an equal span of 100.}
\end{figure}

{\bf Bond dissociation (for acyclic clusters).} To process a bond dissociation, one first samples a bond according to its probability to dissociate. The selected bond locates to a non-leaf node $x$ in a DBT which is identified by its root node $z$. In an acyclic cluster, removal of node $x$ will ultimately split the DBT into two smaller DBTs (we note that the cluster could also dissociate into one free particle and a cluster or into two free particles). If $x$ happens to be the root node $z$, the procedure is trivial. The two child nodes of $z$, $l$ and $r$, simply become root nodes of the two separate DBTs. Otherwise, the final two smaller DBTs are determined by a series of probabilistic decisions, which constitutes a key ingredient of our algorithm. The bond dissociation results in the removal of node $x$ and separates the subcluster into two parts represented by nodes $l$ and $r$. Note that the subcluster represented by node $x$ contributes a site to form the bond at its parent node, $p$, with the other child node of $p$. Therefore, we need to decide at this step which subcluster, $l$ or $r$, provides the site to form bond $p$ and thus will connect to $p$ as a child node. This is done probabilistically. We may assume that the probability of choosing either $l$ or $r$ is proportional to the weight function $g(s_x)$ of the number of free sites contained in the subcluster before the bond $p$ was actually formed. For instance, the number of free sites in subcluster $l$ is $s_l-1$ (in subcluster $r$, the number of free sites is $s_r-1$), where the adjustment $-1$ accounts for the consumption of one site to form bond $x$. The probability of choosing $l$ to connect to $p$ can be then calculated as $g(s_l-1)/(g(s_l-1)+g(s_r-1))$. Without loss of generality, we assume that node $l$ is selected and subcluster $r$ dissociates from cluster $p$. We then update the number of free sites in $p$ as $s_p\leftarrow s_p-(s_r-1)$ and recalculate the weight $g(s_p)$. Applying the same operation, we further decide which of node $r$ and the updated $p$ connects to the parent node of $p$, and so on. This procedure iterates up to the root node $z$ and then in the end obtains two separate DBTs.

Figure~\ref{fig:struct} illustrates that a bond dissociation splits a DBT into two smaller ones in an example ligand-receptor binding system for both acyclic and cyclic clusters. The total number of iterations required to split a DBT equals the depth of the DBT from the dissociating node $x$ to the root node. Each iteration requires generating a random deviate drawn uniformly from the interval $(0,1)$ and updating the weight of the parent node one level up. This bond-breaking procedure is substantially efficient, which has a sublinear cost between $\mathcal{O}(\log{M})$ when a DBT is well balanced and $\mathcal{O}(M)$ when a DBT forms a linear cascade due to sequential attachment of single particles. Both scenarios are rare and unlikely to persist because stochastic bond association and dissociation prevent formation of perpetual linear DBTs or completely balanced DBTs.

{\bf The algorithm.} We now summarize below the algorithm for reversible acyclic aggregation of multisite particles. We note that the memory cost of this algorithm is also substantially reduced in comparison to the site-based algorithm that uses connectivity graphs. The site-based algorithm demands memory to track connections between individual particles and sites, which scales with the total number of sites in a system. The current algorithm has a memory cost scaled by the maximum number of bonds that can be potentially formed in a system, which is usually much less than the total number of sites in the systems. As for other essential memory requirement for simulation, both algorithms maintain lists of bonds and clusters. 

The current algorithm is described as follows:
\begin{enumerate}
\item Initially, sites in every particle are free with no bond formed. The system in the beginning does not have bonds and clusters. Both bond list and cluster list are empty (the cluster list will contain root nodes of DBTs). Initialize the probabilities for bond association and dissociation events.
\item Sample a bond association or a dissociation event based on the probabilities of the two event types at the current step. 
\item If the sampled event is a bond association, create a new bond node as a root node for the DBT of the new cluster, connect the root nodes of the two merging DBTs to the new node, remove the two merging clusters from the cluster list and insert the new cluster into the list. Since a DBT is identified by its root node, one can just remove one root node of the two merging clusters from the cluster list and replace the other with the root node of the new cluster. Insert the new bond into the bond list. Go to Step 5.
\item If the sampled event is a bond dissociation, identify an individual bond to break by searching over the bond list. Split the DBT to which the chosen bond locates into two smaller DBTs. Replace the root node of the splitting cluster in the cluster list with one root node of the two new smaller clusters and insert the other root node into the list. Remove the dissociated bond from the bond list.
\item Update the probabilities for bond association and dissociation events.
\item Repeat Step 2 until the simulation stops.
\end{enumerate}

The above procedure does not explicitly include tracking time evolution in simulation, which can be included as shown in our application below. A simulation of a system starts out with all free particles without bonds, and after a transient dynamics the system will eventually relax to its equilibrium where the rate of bond formation is balanced by the rate of bond breaking.  As we will show, this algorithm may provide a substantial speedup for processing bond dissociations in high density clusters.

The procedure for simulating cyclic aggregation that allows loop formations in clusters is largely the same as described above with some modifications to handle cyclic bonds. As mentioned in the previous section, processing a cyclic bond formation is also trivial. Whenever an intracluster site pair forms a bond, a new node is created with only one subtree that corresponds to the same cluster contributing both binding sites. For a cyclic cluster, breaking a bond may or may not split the corresponding DBT into two smaller ones. If the breaking bond $x$ happens to only have a single child, we connect this child node directly to the parent node of $x$ and update the weights of all subsequent parent nodes up to the root node of the DBT. In this case, because the bond $x$ is part of a loop in the cluster, its dissociation does not split the cluster into two. If node $x$ has two child nodes, the procedure is identical to that of splitting an acyclic DBT until the iteration meets one upper level parent node $p$ that has only one child. In such a case, we need to determine how the two subclusters contribute a pair of sites to form the bond at node $p$. There are two possible ways, to be determined probabilistically: (1) One of the two subclusters contributes both sites. This subcluster will then connect to $p$ as a single child node and the other subcluster remains separate for further processing to the upper level of the DBT. (2) Each subcluster contributes a site. In this case, the two subclusters connect to node $p$ as two children nodes, and no further probabilistic decisions are needed except for updating the weights of all the upper level parent nodes up to the root node.

Figure~\ref{fig:struct} illustrates how to process cluster aggregation using DBTs for an example system of trivalent ligands with three binding sites aggregate with bivalent receptors with two binding sites (TLBR model). Especially, we note that an equivalent class of DBTs exists for each cluster with a distinct connectivity, and vice versa. It is possible,  by probabilistic mapping, to systematically convert a DBT into an ensemble of cluster connectivity graphs with same numbers of particles and bonds for inspection. For instance, Fig.~\ref{fig:struct} shows that cluster I may be represented as DBT I or II depending on the sequence of bond formation. The stochasticity in breaking a bond in a cluster can also result in diverse fragmentation scenarios (Fig.~\ref{fig:struct} right panel).

\section{Application}\label{sec:app}

To demonstrate our algorithm and compare its performance to the conventional site-based algorithm, in this section we specialize to simulate aggregation in the TLBR system. The system is representative to aggregation of a mixture of heterogeneous particles with multiple complementary binding sites and thus serves as a benchmarking system for comparison between the current method and the conventional one that uses graph traversals. The acyclic TLBR aggregation was originally studied analytically by Goldstein and Perelson~\cite{goldstein1984equilibrium} who used an equilibrium model to obtain cluster size distribution and showed the existence of sol-gel phase transition under a certain range of parameter values. Results from a recent Monte Carlo simulation study~\cite{yang2008kinetic} showed agreement with Goldstein and Perelson's equilibrium theory. 

In the TLBR system, a population of extracellular ligands, each of which has three identical binding sites, interact with a population of receptors distributed on the cell surface, each of which has two identical binding sites. A bond can be formed only between a ligand site and a receptor site. The TLBR system distinguishes two kinds of bond formations: (1) Free ligands are first recruited to cell-surface receptors and (2) bound ligands with free binding sites can subsequently crosslink receptors and induce receptor aggregation. A ligand-receptor bond can dissociate spontaneously. We apply the law of mass action to account for the rates of bond association and dissociation. Here, we simply assume that the probability for a cluster to contribute a receptor (ligand) site is proportional to the number of free receptor (ligand) sites in the cluster, i.e., $g(s_c)\equiv s_c$. 

The system involves three rate processes parameterized by different rate constants: (1) Free ligands precipitating to bind cell surface receptors with a rate constant $k_+$. The rate can be calculated as:
\begin{equation}
r_1=k_+v_lF_L(v_rN_R-N_B) \ , 
\end{equation}
where $v_l$ and $v_r$ are valences (number of binding sites) in the ligand and the receptor molecule, respectively. $F_L$ is the number of free ligands in solution, $N_R$ and $N_B$ are the numbers of total receptors and bonds, respectively. (2) Receptor crosslinking by ligands already bound to receptors with a rate constant $k_{++}$. The rate is calculated as:
 \begin{equation}
 r_2=k_{++}((v_l(N_L-F_L)-N_B)(v_rN_R-N_B)-(1-\phi)Z) \ , 
 \end{equation}
 where $\phi$ is a parameter that characterizes the average probability of an intracluster site pair to form a bond and takes a value within the interval $[0,1]$. The aggregation becomes acyclic when $\phi=0$. The term $(v_l(N_L-F_L)-N_B)(v_rN_R-N_B)$ accounts for the total product between free ligand sites and receptor sites on the cell surface and the term $(1-\phi)Z$ accounts for a reduction by the production of intracluster ligand and receptor sites adjusted by the parameter $\phi$. The quantity $Z$ is the sum of intracluster site combinations over all clusters, $Z=\sum_{i=1}^{N_C}l_ir_i$. $l_i$ and $r_i$ are the numbers of free ligand and receptor sites in cluster $i$. respectively. In practice, since each event only affects a smaller number of clusters, $Z$ can be calculated after each event by iterative update to avoid summing over the entire array of clusters~\cite{yang2008rejection}. And, (3) Ligand-receptor bond dissociation with a rate constant $k_{\rm off}$. The rate is proportional to the number of bonds and is calculated as:
 \begin{equation}
r_3=k_{\rm off}N_B \ .
\end{equation}
Evaluation of the rates for the above processes takes a constant cost for each event if quantities including $F_L$ and $N_B$ are bookkept during simulation. 

\begin{figure}[ht]
\centering
\includegraphics[scale=0.5]{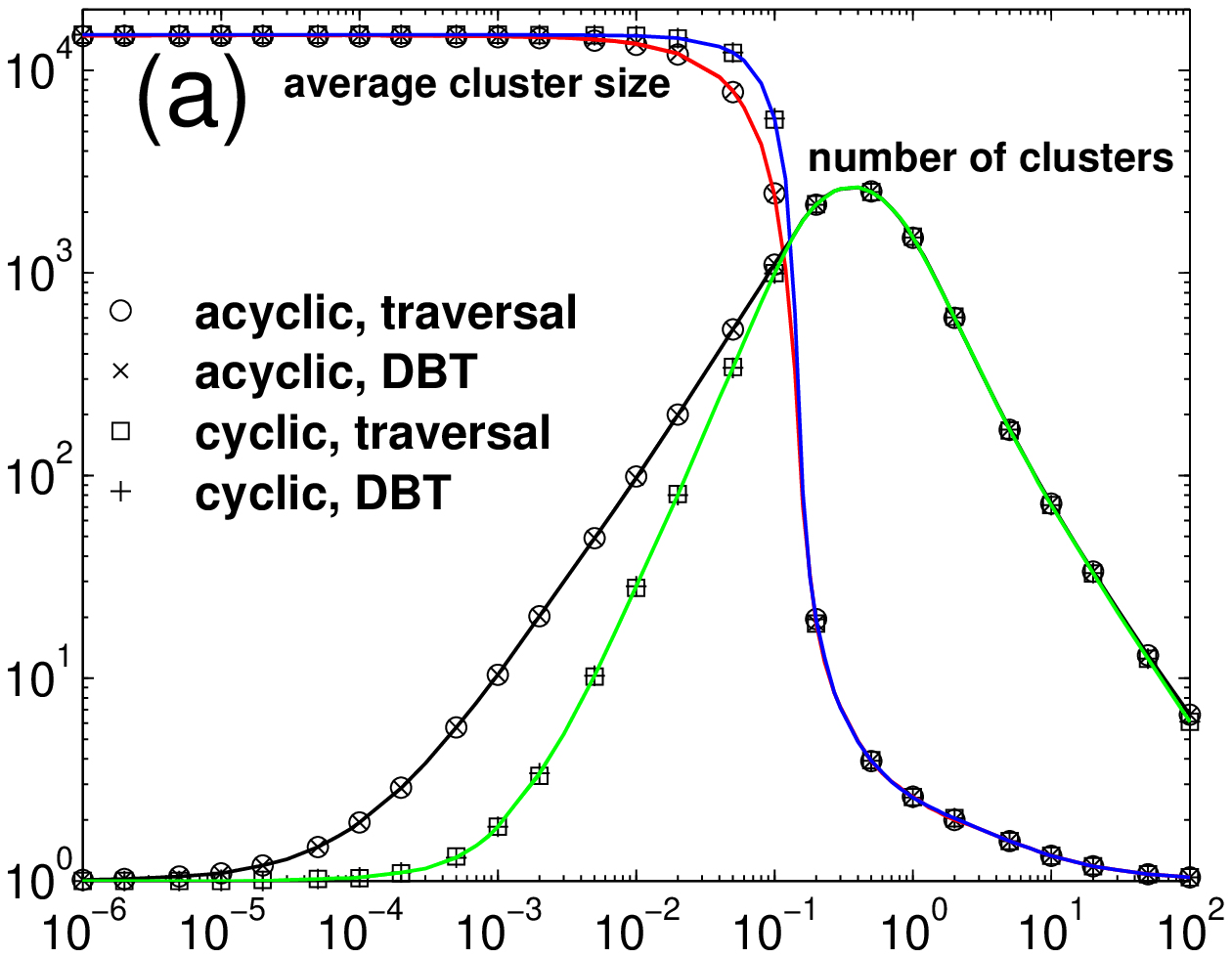}

\vspace{-0.1in}

\includegraphics[scale=0.5]{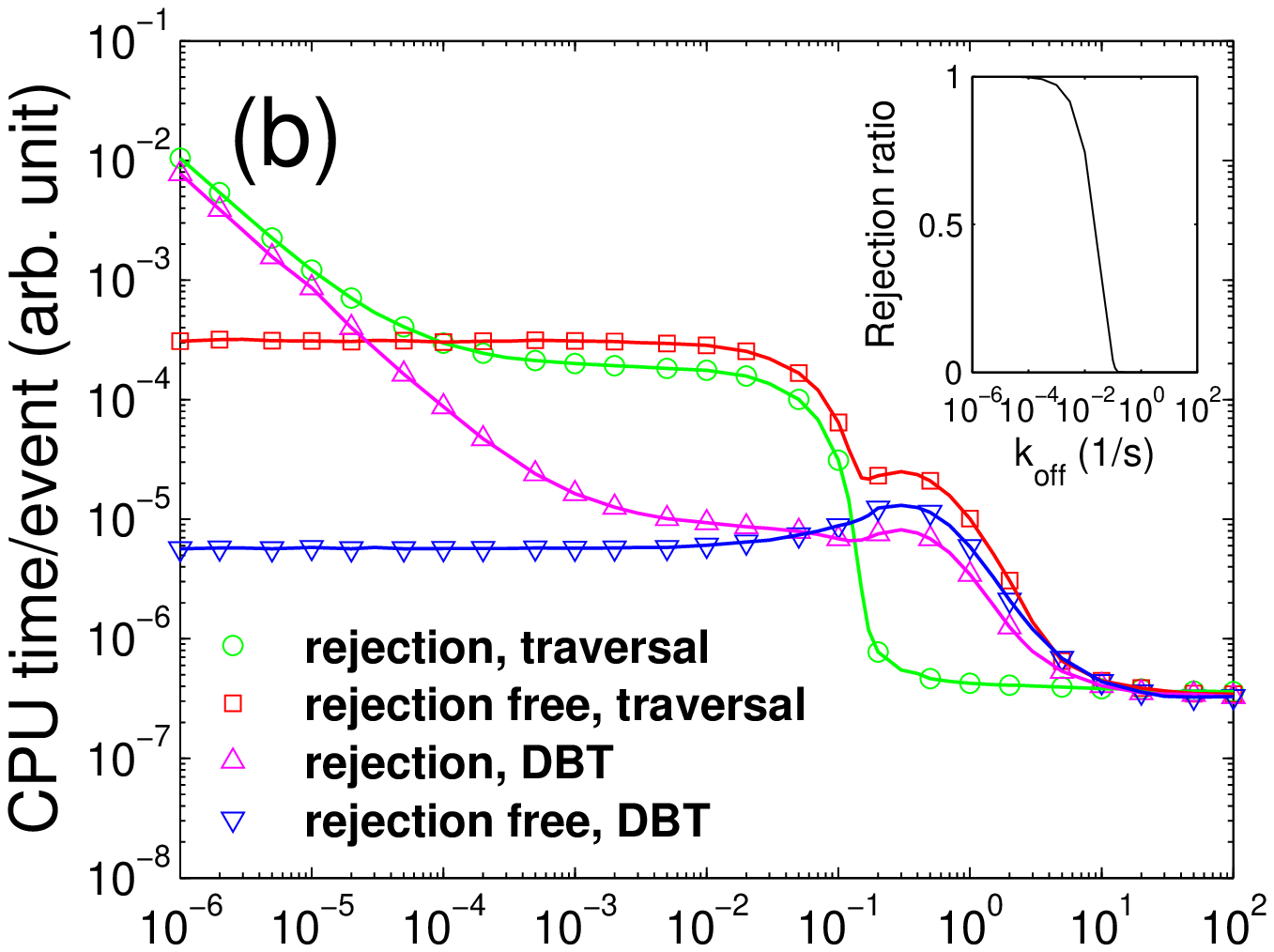}

\vspace{-0.1in}

\includegraphics[scale=0.5]{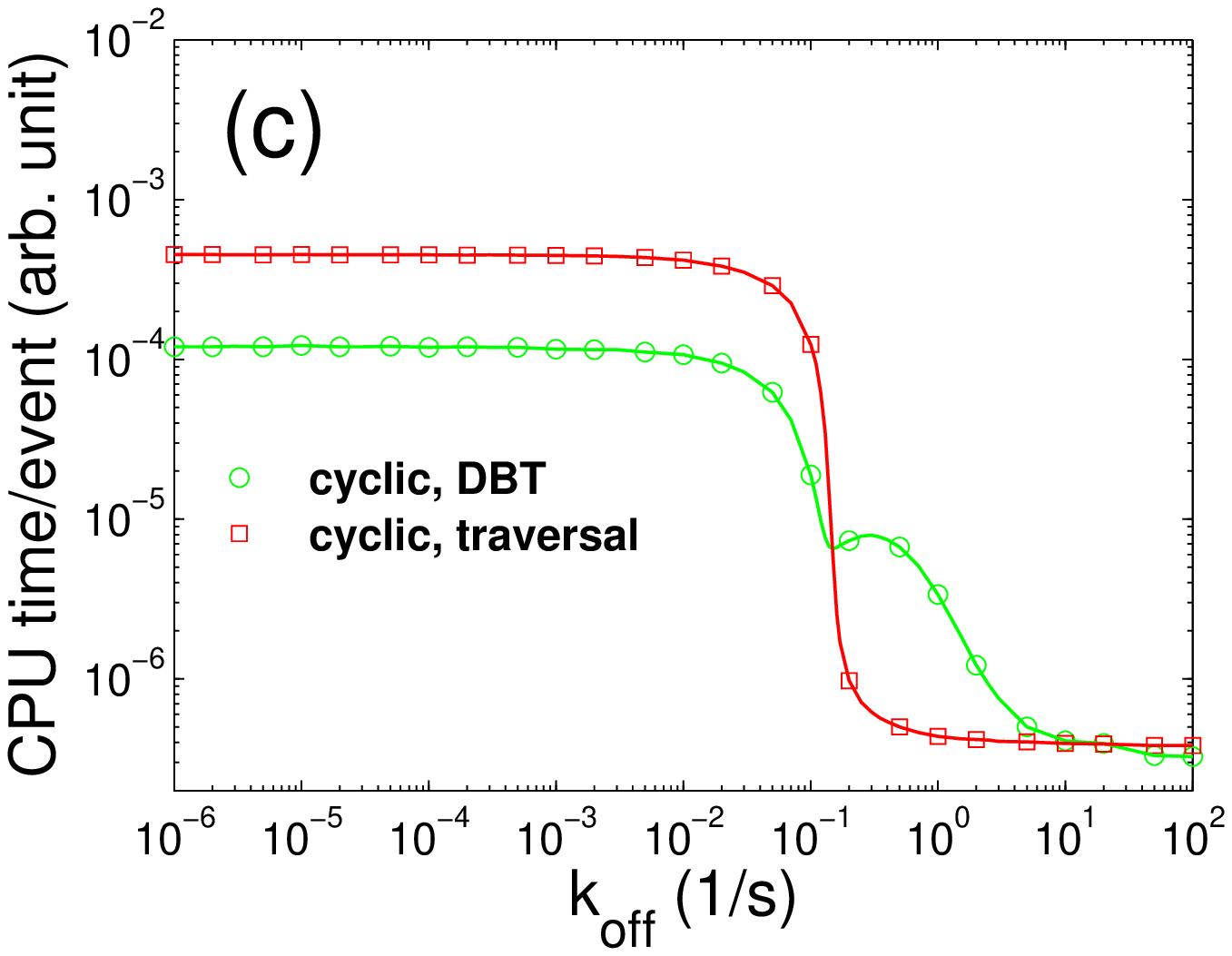}
\caption{\label{fig:cpu} (Color online) (a) Average cluster size and the number of clusters, by methods using DBTs or graph traversals for acyclic and cyclic aggregations. The results were obtained by averaging 5000 samples (each sample was separated by 100 events) at the equilibrium. (b) Performance of four schemes for simulating acyclic aggregation of the TLBR system: rejection or rejection-free sampling with or without employing DBTs. Inset: rejection ratio [the ratio of the number of effective events to the number of all events] in different phase regimes. The mean CPU time per event was obtained by averaging after the system reached the equilibrium. (c) Performance comparison between DBT and graph traversal methods for simulating cyclic aggregation ($\phi=1$). Parameters are identical to the ones indicated in Fig.~\ref{fig:dep}.}
\end{figure}

The simulation follows the typical procedure of kinetic Monte Carlo or stochastic simulation algorithm~\cite{bortz1975new,gillespie1977exact,blue1995faster}. At the start of a simulation, all ligand and receptor sites are free with no bond formed. For each iteration, to record the lapse of time, we first determine the waiting time $\tau$ for the next event, which is sampled from an exponential distribution with the mean waiting time $\langle\tau\rangle=1/r_{\rm tot}$, where $r_{\rm tot}=r_1+r_2+r_3$. We then select a rate process that fires the next reaction event. The probability of a process to be selected is proportional to its rate. Finally, we update the configuration of the system and recalculate the reaction rates.

Figure~\ref{fig:dep} shows that the DBT depth in acyclic aggregation has a very slow growth against the increase in the cluster size. The growth of the DBT depth can be fit to a monomial function of the number of bonds as $M^{0.45}$. Cyclic aggregation exhibits a steeper growth of the DBT depth against the cluster size because forming intracluster bonds increases the DBT depth on top of an acyclic cluster with the same number of receptors and ligands. The deepest cyclic DBTs that correspond to the largest clusters, on average has the depth only one tenth (about 4,000 vs. 50,000) of the cluster size measured by the number of particles and bonds in the cluster. For the cyclic aggregation shown in Fig.~\ref{fig:dep}, we allow each free ligand-receptor site pairs (both intracluster and intercluster site pairs) has an equal probability to form a bond. Note that this assumption overestimates the probability of both intracluster and intercluster bond formation because geometric constraints may prohibit interactions between certain intracluster site pairs~\cite{monine2010modeling}. For more realistic models, one can use a different function $g(s)$ to account for the probability of a cluster to provide binding site. Upon each association event, a trial intracluster ligand-receptor site pair is accepted to form a bond with a probability $\phi$. Here by setting $\phi=1$, we intend to present a worst-case scenario for cyclic aggregations in terms of the average DBT depth and expect that the performance of simulating any intermediate cyclic aggregation model ($0<\phi<1$) of the TLBR system will lie between this extreme model and the one of acyclic aggregation ($\phi=0$). 

Simulation results verified that for both acyclic and cyclic aggregations, our algorithm is statistically identical to the site-based graph traversal method in obtaining the average cluster size and the number of clusters $N_C$ under varying dissociation rate constant $k_{\rm off}$ (see Fig.~\ref{fig:cpu}(a)). The cluster size is measured by the number of receptors in a cluster (not including free receptors). The average cluster size is given by: $\sum_{n=1}^{N_R}n^2x_n/(N_R-F_R)$, where $x_n$ is the number of clusters of size $n$, $F_R$ is the number of free receptors. The result shows that cyclic aggregation produces less clusters and bigger average cluster size than acyclic aggregation within the middle range of $k_{\rm off}$ (between $10^{-4} s^{-1}$ and $10^{-1}s^{-1}$) and the two types of aggregation converge at both high density region and weakly aggregated region. The number of clusters $N_C$ reaches a maximum about 3000 at $k_{\rm off}=0.4 s^{-1}$, which is much lower than the total number of receptors $N_R$ and falls down sharply as the average cluster size reaches maximum.

To simulate acyclic aggregation, rejection sampling can be employed to enforce the loopless condition required for receptor crosslinking. Upon each crosslinking event, the rejection sampling in the site-based algorithm randomly picks out  a pair of sites. The algorithm rejects the trial sites if the two sites are found in a same cluster. Otherwise, the sites are accepted to form a bond. In contrast, the current algorithm using DBTs processes crosslinking by picking out two clusters $x$ and $y$ for ligand and receptor sites, respectively, based on the number of free binding sites in the clusters. The event is rejected if clusters $x$ and $y$ are identical ($x=y$). However, rejections can slow down a simulation when the system has a high density cluster that contains a large number of particles. One can measure the extent of rejection sampling using a rejection ratio, $\theta$, defined as the probability that a pair of sampled sites are rejected for binding. For example, the instantaneous rejection ratio can be calculated for the acyclic TLBR system as: $\theta= {k_{++}Z}/r_{\rm tot}$. In practice, we can obtain the average rejection ratio in the steady state as rejected events normalized by the total number of events including null events. As shown in Fig.~\ref{fig:cpu}(b) for an algorithm using either site-based connectivity graphs or DBTs, efficiency of simulation decreases when rejected samples become dominant in the rejection sampling (e.g., $\theta>0.9$). 

To simulate an acyclic aggregation system with high density clusters, a rejection-free sampling~\cite{yang2008rejection} of binding sites is required to overcome the bottleneck caused by a high rejection ratio in the rejection sampling that excludes intracluster site pairs from binding.  Fig.~\ref{fig:cpu}(b) shows the performance comparison between different methods that use graph traversals or DBTs with or without rejection-free sampling. The combination of DBTs and the rejection-free sampling is superior to other approaches. Except only for the method using rejection sampling with graph traversals, a hump (about $k_{\rm off}=0.4 s^{-1}$) in each curve in Fig.~\ref{fig:cpu}(b) for simulating acyclic aggregation and the DBT method in Fig.~\ref{fig:cpu}(c) for simulating cyclic aggregation reflected a small performance penalty due to sampling over a maximum number of clusters for binding clusters near the phase transition boundary. For simulations of cyclic aggregation, the performance comparison between the DBT and site-based methods is shown in Fig.~\ref{fig:cpu}(c). In this fully cyclic aggregation model ($\phi=1$) because every free ligand-receptor site pair is allowed to crosslink on the cell surface, no rejection sampling is required in simulations.  At the high density region ($k_{\rm off}<0.01$), the method using DBTs is four times faster than the site-based algorithm using graph traversals.

\section{Conclusion}

We have presented an efficient kinetic Monte Carlo procedure for simulating reversible aggregation of multisite particles, especially for systems with a large number of particles that nucleate into high density clusters. The algorithm generates statistically identical results as the standard method with considerably less time and space complexity in computation. To avoid costly operations of traversals of cluster connectivity graphs, the algorithm records clusters and processes bond formation and breaking using  dynamic bond trees that track the hierarchy of cluster aggregation. The method provides a fast means to evaluate aggregation of multisite particles and can be in general adapted to simulate a wide class of particle aggregation and self-assembly.  

In this paper, we demonstrated the scaling of our algorithm against cluster size with the TLBR system that is prototypical to more general reversible aggregation of multisite particles. We note that the standard graph traversal algorithm has a complexity of $\mathcal{O}(M+N)$ in processing bond breaking, which scales linearly with the number of particles and bonds in an aggregate. In comparison, the worst-case scenario in the current method using DBTs has a complexity of $\mathcal{O}(M)$, which scales linearly only with the number of bonds in an aggregate and in principle improves over the standard method. This worst case is however probabilistically unlike to persist for any typical system such that the complexity of our algorithm is always below $\mathcal{O}(M)$. For example, $\mathcal{O}(M)$ complexity requires the DBTs used for storing an aggregate has a height that is equal to the number of bonds so that the DBT updating takes $\mathcal{O}(M)$ time after a bond was randomly sampled to break in the original DBT. For this situation to persevere, it requires that each acyclic bond association and dissociation must happen between a single particle and an aggregate (or another single particle) to maintain each DBT as a linear chain with no branches, which effectively precludes aggregation between multiparticle clusters. On the other hand, the scaling of $\mathcal{O}(\log{M})$ requires well-balanced DBTs (i.e., balanced binary trees), which implies each bond association must happen between two clusters with identical DBTs and each bond dissociation must result in two identical clusters. This scenario obviously can only happens fortuitously during a kinetic Monte Carlo simulation. The actual performance of the algorithm will depend on the property of a particular aggregation system itself and specific parameters used in a simulation including the valence on particles, kinetic rate constants, interaction rules, and etc.

\section{Acknowledgement}

We thank William Hlavacek and John Pearson for helpful discussions. This work was supported by National Science Foundation of China through grant 30870477 (J.Y.).

%

\end{document}